\documentstyle[11pt,epsfig,epsf]{article}
\def\lq{\left [}
\def\rq{\right ]}
\newcommand{\qv}{|\vec q|}
\newcommand{\be}{\begin{equation}}
\newcommand{\ee}{\end{equation}}
\newcommand{\bea}{\begin{eqnarray}}

\newcommand{\eea}{\end{eqnarray}}
\newcommand{\nn}{\nonumber}
\newcommand{\dd}{\displaystyle}
\newcommand{\spur}[1]{\not\! #1 \,}
\begin{document}

\title{\hfill 
$\mbox{\small{\begin{tabular}{r}
${\textstyle Bari-TH/98-307}$
\end{tabular}}}$ \\[1truecm]
Semileptonic and Rare $B$-Meson Transitions\\  
in a QCD Relativistic Potential Model}
\author{P. Colangelo$^a$, F. De Fazio$^a$, M. Ladisa$^{a,b}$,\\ 
G. Nardulli$^{a,b}$, P. Santorelli$^{a,b}$ and A. Tricarico$^b$} 
\maketitle

\begin{it}
\begin{center}
$^a$Istituto Nazionale di Fisica Nucleare, Sezione di Bari, Italy\\
$^b$Dipartimento di Fisica dell'Universit\`a di Bari, Italy
\end{center}
\end{it}

\begin{abstract}
Using a QCD relativistic potential model, previously applied to the
calculation of the heavy meson leptonic constants,
we evaluate the form factors governing the exclusive decays 
$B\to\rho\ell\nu$, $B\to K^*\gamma$ and
$B\to K^*\ell^+\ell^-$.
In our approach the heavy meson is
described as a $Q\overline q$ bound state, whose wave function is
solution of the relativistic Salpeter equation, with an instantaneous
potential displaying Coulombic behaviour at small distances and linear
behaviour at large distances. The light vector meson is described by using
a vector current interpolating field, according to the Vector
Meson Dominance assumption. A Pauli-Villars regularized propagator is assumed 
for the quarks not constituting the heavy meson.
Our procedure allows to avoid the description of the light meson in terms of
wave function and constituent quarks, and consequently the problem
of boosting the light meson wave function. 

Assuming as an input the
experimental results on $B\to K^*\gamma$, we evaluate all
the form factors describing the $B\to \rho, K^*$ semileptonic and rare
transitions. The overall comparison with the data,
whenever available, is satisfactory. 
\end{abstract}
\newpage

\section{Introduction}
\label{s:1}

$B$ meson decays play a central role in particle physics, as witnessed
by the considerable amount of experimental data collected, mainly at CESR, LEP, 
Tevatron and SLAC accelerators. More of all,
the importance of $B$ decay processes is related to
 the results which will come in the near future
from BaBar and Belle experiments at the dedicated SLAC and KEK
$B$-facilities, and from the LHC-B experiment  at CERN, whose 
main goal is
the analysis of $CP$ violation in the $B$ system \cite{physbook}.
Consequently, theoretical efforts are greatly projected towards the
determination of methods to extract, from the data, the elements of the
Cabibbo-Kobayashi-Maskawa matrix ($CKM$), since the complex nature of
this matrix is the source of $CP$ violation within the Standard Model
($SM$). From this point of view, heavy-to-light decays are of prime
interest, since $b \to u$ transitions offer the possibility of determining the
poorly known matrix element $V_{ub}$, while $b \to s$ processes,
forbidden at tree level in $SM$, give access to $V_{ts}$ and, in addition, 
represent a powerful tool to
investigate possible effects of physics beyond $SM$. 

In this paper we address both $b \to u$ and $b \to s$-induced decay channels, 
and in particular we analyze the exclusive semileptonic and rare
$B$ decays to a  nonstrange $\rho$ and strange $K^*$ vector mesons.

The branching ratio of the
semileptonic decay $B \to \rho \ell \nu$ has been
recently measured by the  CLEO collaboration \cite{cleo96}: 
\be
{\cal B}(B^0 \to \rho^- \ell^+ \nu)=(2.5 \pm 0.4^{+0.5}_{-0.7}\pm0.5) \;
\times \; 10^{-4} \;\;\;. \label{cleorho}
\ee
\noindent 
From the experimental viewpoint, due to the overwhelming $b \to c$ 
transitions, such
decay is not easily accessible, and one has to select leptons in a high
momentum range which can be reached in the $b \to u \ell \nu$
transitions, but not in the  $b \to c \ell \nu$ process. In addition to
such experimental difficulty, in order to extract $V_{ub}$
from the data one has to deal with the theoretical uncertainty arising
from the evaluation of the hadronic $B\to \rho$ matrix element. 

The analysis of the rare decays $B \to K^* \gamma$ and $B \to K^* \ell^+ 
\ell^-$  presents analogous 
uncertainties. Experimental data already exist for both the
inclusive $b \to s \gamma$ and the exclusive decays with a real photon in
the final state: 
\bea
{\cal B}(b \to s \gamma)&=&(2.32 \pm 0.57\pm0.35)\; \times \;10^{-4}
\cite{cleo95} \label{cleorari} \\ 
{\cal B}({\bar B}^0\to  K^{*0} \gamma)&=&(4.0\pm1.7\pm0.8)
\; \times \; 10^{-5} \cite{cleo93}
\label{cleorari1}  \\
{\cal B}(B^-\to K^{*-} \gamma)&=&(5.7\pm3.1\pm1.1)\; \times \;10^{-5}~.
\cite{cleo93} \nonumber
\eea
\noindent  
These results constrain the parameters featuring various new physics
models, since rare $B$ decays are particularly sensitive to
effects beyond SM, 
but also in this case the interpretation depends on the reliable
evaluation of the relevant hadronic $B \to K^*$ matrix elements.

To deal with  such non perturbative quantities of the $B$
physics few theoretical approaches are available so far. 
Directly related to the QCD description of strong interactions are
Lattice QCD and QCD Sum Rules.
Lattice QCD, based on the
procedure of discretizing the space-time, allows a numerical evaluation
of the hadronic matrix elements \cite{latticerev}.
QCD Sum Rules, in
the versions of three-point Sum Rules (SR) and Light Cone Sum Rules
(LCSR) are based on fundamental
properties of quark current correlators, such as the analiticity and 
the possibility of expanding at short-distances or on the light-cone.
\cite{sumrev}.
The above approaches have their own advantages and drawbacks, 
in particular systematic uncertainties related to the quenched approximation 
for Lattice QCD, and errors induced by the truncation of the Operator
Product Expansion for the Sum Rules. Therefore, it is
worth looking for other approaches that, while being less 
fundamental, present nevertheless the advantage of computational simplicity
and offer at the same time a sufficiently deep physical insight. 

To study the transitions between the $B$ meson and the light
vector mesons $\rho$ and $K^*$ and  
to compute the relevant hadronic matrix elements
we employ in this paper a QCD relativistic potential model. The
main aspect of the model is that the heavy meson is
described as a $Q\overline q$ bound state; the wave function is
obtained by solving
the relativistic Salpeter equation \cite{salpeter}, with an
instantaneous potential displaying Coulombic behaviour at short
distances and linear (confining) behaviour at large distances.
The light vector meson is described by using a
 vector current interpolating field, following 
the Vector Meson Dominance Model. Finally, a Pauli-Villars regularized 
propagator is assumed for 
the quarks  not constituting the heavy meson.
Since we do not describe the light meson in terms
of wave function and constituent quarks, we can avoid the problem of 
boosting the wave function, a point which is, in general, a source of
considerable ambiguity. 

Our approach represents an extension to the problem of
heavy-light transitions of the work in Ref. \cite{piet}
where the spectrum of $\bar q Q$ mesons and the leptonic constants, both for 
finite heavy quark masses and in the infinite limit are analyzed. We 
begin by describing the heavy meson wave equation in Section \ref{s:2}, and the
interaction with the hadronic current in Section \ref{s:3}. 
In Section \ref{s:4} we apply
the model to the evaluation of the leptonic decay constant
$f_B$ with the aim of showing that, in the infinite heavy quark mass limit, the
results of the method presented in this work
coincide with those of Ref. \cite{piet}. 
The model has one free parameter: the mass of the shape function describing 
the deviation from the free propagation of one of the light quarks.
This parameter is fixed in Section \ref{s:5} by fitting the experimental
results in Eq. (\ref{cleorari}). As a consequence, we are able to calculate
all the form factors describing the $B\to V$ transitions. In Section
\ref{s:6} we compute the decay width $B\to\rho\ell\nu$ and compare our
outcome  with the results of other non perturbative approaches, as well as with
the experimental data. In the Appendix we collect the relevant
formulae for the various form factors describing the heavy-to-light 
transitions.

\section{Heavy meson wave function}
\label{s:2}

As discussed in the Introduction, an important aspect of the model is
provided by the heavy meson wave function  arising from a constituent
quark picture of the heavy hadron. The heavy meson $H$
is described as a bound state of two constituent quarks: a heavy (Q)
quark and a light ($\bar q$) antiquark. Denoting by $\vec k$ the 3-momentum
of the quark $Q$ in the meson rest frame 
($- \vec k$ is the antiquark momentum), the
momentum distribution of the constituent quarks is provided by the
wave function $\psi(\vec k)=\psi(k)$, whose Fourier transform
$\Psi(\vec r)$ is solution of the Salpeter equation \cite{salpeter}
\be
\left [ \sqrt{ -\nabla^2 + m^2_Q} + \sqrt{ -\nabla^2 + m^2_q}
+V(r)\right ] \Psi(\vec r)=m_H \Psi(\vec r)~.\label{1}
\ee
The variable $r$ in (\ref{1})
is the interquark distance, $m_H$ is the heavy hadron mass and
$V(r)$, as discussed in \cite{piet,cea}, is given by the
Richardson potential \cite{rich}:
\be
V(r)=\frac{8 \pi}{33-2 n_f} \Lambda \left (\Lambda r -\frac{f(\Lambda
r)}{\Lambda r}\right )~~~~~~~(r\geq r_m) \label{2}
\ee
with the function $f(t)$ given by
\be
f(t)=\frac{4}{\pi}\int_0^\infty
dq \frac{sin(q t)}{q}\left [ \frac{1}{ln(1+q^2)}-\frac{1}{q^2}\right
]~.\label{3}
\ee
As shown by eqs.(\ref{2}),(\ref{3}), the Richardson potential 
increases linearly at large distances, 
thus providing confinement of the quarks, whereas 
at short distances it displays a Coulombic behaviour with running $\alpha_s$, 
as dictated by perturbative QCD.
For $r<r_m=\frac{4 \pi \lambda}{3 m_H}$ ($\lambda$ is a parameter to be fitted
within the model) we assume in
Eqs. (\ref{2}) and (\ref{3})  $V(r)=V(r_m)$. The reason for
this cut-off of the potential is in the fact that (\ref{2}), 
for $r\to 0$, exhibits a Coulombic divergence. Such a divergence is harmless 
in the nonrelativistic Schroedinger equation; on the other hand, 
if one of the quark masses ($m_q$) is light and the
relativistic kinematics, embodied in the Salpeter equation, is
adopted, the Coulombic divergence of the potential produces an unphysical
logarithmic divergence of the wave function at the origin \cite{durand}.
The form of the modified Richardson potential can be fixed by studying the
problem of quark-hadron duality in $e^+ e^- \to hadrons$ \cite{cea}.
The potential (\ref{2}) does not include spin terms, and therefore, 
the $J^P=0^-,~1^- ~Q\overline q$ mesons are degenerate in mass,
an approximation which is expected to work better in the limit
$m_Q\to\infty$. 

Notice that, due to the simple choice of the interquark potential in 
Eq. (\ref{2}), we do not try to apply the wave equation to mesons
comprising only light $(u,d,s)$ quarks. As a matter of fact, in such a
case the interaction between the quarks cannot be described by the
simple form (\ref{2}), since the spin terms cannot be neglected; moreover
the assumption of the constituent quark picture and the instantaneous
interaction is more dubious for light mesons. For example, for the
description of the light pseudoscalar meson octet, the notion of light
pseudoscalar particles as Nambu-Goldstone bosons has to be implemented.
For this reason we describe the light mesons 
by using effective fields, in the spirit of the chiral effective theories and
the Vector Meson Dominance Model.
\par
For $\ell = 0$ (S-wave) heavy mesons, the only ones of interest here,
Eq. (\ref{1}) reduces to
\be
[V(r)-m_H]\Psi(r)+ \frac{2}{\pi r}\int_0^\infty dr^\prime \Psi(r^\prime)
r^\prime
 \int_0^\infty dk\left [ \sqrt{k^2+m_Q^2}+ \sqrt{k^2+m_q^2}\right]
sin(kr)~sin(kr^\prime)\;=\;0\;\;\;.
\ee 
It can be solved by a numerical procedure, fitting the parameters of the model
in order to reproduce the experimental meson spectrum.
The following values for
the parameters are obtained: $\Lambda=397$ MeV, $\lambda=0.6,\,m_u=m_d=38$ MeV,
$m_s=115$ MeV, $m_c= 1452$ MeV, $m_b=4890$ MeV. The resulting fit of the
heavy meson masses can be found in Ref. \cite{piet}.

The B meson reduced wave function $u(k)$ can be defined  as follows:
\be
u(k)=k \psi(k)
\ee
with the function
\be
\psi(k)= \int_0^\infty r~\Psi(r)~sin(kr)~dr
\ee
normalized following the relativistic prescription
\be
\int \frac{d^3 k}{(2\pi)^3} |\psi(k)|^2=2 m_B~.
\label{norm}
\ee

\begin{figure}[ht]
\centerline{\epsfig{file=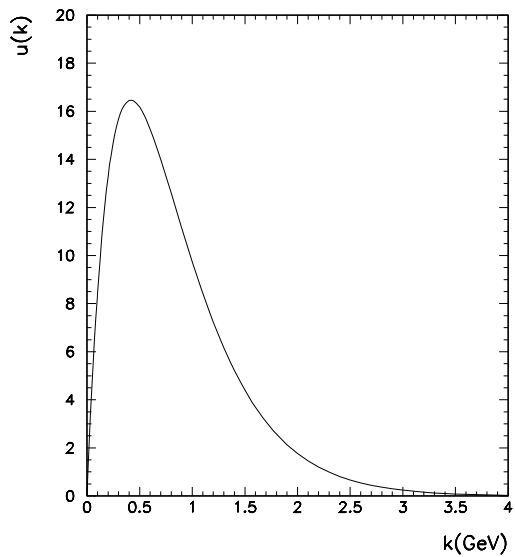,height=8cm}}
\caption{The $B$ meson reduced wave function $u(k)$.}
\label{f:fig1} 
\end{figure}

An analytical representation of the $B$ wave function, which fits
the numerical solution obtained by the Multhopp method \cite{multhopp},
is given by
\be
u(k)=4\pi\sqrt{m_B \alpha^3}\;k\; e^{-\alpha k}
\label{wf}
\ee
with $\alpha=2.4$ GeV$^{-1}$, as plotted in Fig. \ref{f:fig1}.

\section{Interaction with the hadronic currents}
\label{s:3}

Let us consider the matrix element:
\be
<V|J^\mu|H>\label{8}
\ee
where $H$ is a heavy ($Q\bar q$) meson and $J^\mu$ is a hadronic
current:
\be
{\bar {q^\prime}} \Gamma^\mu Q~. \label{9}
\ee
In Eq. (\ref{9}), $q^\prime$ is a light quark, $\Gamma^\mu$ a vector
combination of $\gamma$ matrices and momenta and $V$ an hadronic state
not containing heavy quarks. In the sequel we shall consider only the
case where $H$ is the $B$ meson with $J^P=0^-$ and $V=$ hadronic vacuum
(in this case $q=q^\prime$) or  a $q^\prime\bar q$ light vector meson.
The formalism can be immediately extended to other cases, such as
transitions between $B$ and a light pseudoscalar meson, transitions
between two heavy mesons with a current containing both heavy quarks or 
light quarks, etc., but we defer a detailed treatment of all these cases
to a future publication. 
\par
In the constituent quark model, the straightforward approach to the
evaluation of the matrix element (\ref{8}) would be as follows. The
heavy meson state $|H>$ is decomposed on a limited Fock base, containing
only $Q\bar q$ pairs, with a momentum distribution weighted by
$\psi(k)$. Since one knows the meson wave function only in the rest
frame, the matrix element (\ref{8}) has to be evaluated at rest. In
general this is  a limitation which allows to compute only the
vacuum-single particle matrix element, while, for two single particle
states, the form factors can only be evaluated at zero recoil. As a
matter of fact, at zero recoil both particles are at rest and the form
factors describing (\ref{8}) can be extracted in the meson rest frame by
working out the product of quark and antiquark operators appearing in
the states $|H>,~|V>$ and in the current (\ref{9}). This procedure has
been applied in \cite{piet} to the evaluation of the matrix element 
\be
<0|\bar q \gamma^\mu \gamma_5 b|\bar B(p)>=if_B p^\mu\label{fb}
\ee
and in \cite{tedesco} to the calculation of the matrix element
$<D^{(*)}|J^\mu|B>$
at zero recoil.
\par
Such standard procedure cannot be immediately applied to the evaluation
of the transition matrix element $B\to V$ ($V$ is a light vector meson)
for
general values of the momentum transfer $q^2$ for  several resons:
1) as discussed above, the quark constituent picture and the
approximation of the instantaneous interaction are too crude for light
mesons; 2) even in the approximation of the instantaneous interaction,
the potential $V(r)$ in $(2)$ and $(3)$ is unrealistic for light mesons,
since one is dropping spin terms that are not negligible for the light
degrees of freedom; 3) to obtain the different form factors at various
values of $q^2$, one would need a reliable method to boost the wave
functions in a moving frame: while some recipes are available (see e.g.
\cite{thomas}) the prescription is not unique because our approach, similarly 
to all the potential
models with instantaneous interaction, does not exhibit full relativistic
invariance. 
\par
Instead of following this approach we propose to consider the following 
representation:
\be
<V(p^\prime, \epsilon^*)|J^\mu|B(p)>\simeq 
{m_V^2 \over f_V} \epsilon^{*\nu} \int dx \; e^{i p^\prime x}
<0|T(V_\nu(x) J^\mu (0) ) |B(p)> \label{correl} 
\ee
where  $V_\nu={\bar q} \gamma_\nu q^\prime$ 
with  ${ q^\prime},  q$ 
light quarks (=$u,d,s$), $m_V$ is the vector
meson mass
and $f_V$ is a coupling thatcan be computed  by the decay widths.
 The approximation (\ref{correl}) can be seen as the implementation of the 
Vector Meson Dominance Model, in the limit $p^{\prime2} \to 0$. 
The calculation of (\ref{correl}) would follow from the usual Feynman rules, 
with some important modifications suggested by the confined nature of the quark 
belonging to the heavy and light mesons. They can be summarized as follows.
\par\noindent
1) The light meson is described by an effective field operator, in the
spirit of the chiral effective field theories. To describe the light 
vector $1^-$ state, we introduce the effective operator 
\be
\Phi^\mu=\frac{m^2_V}{f_V} \;\;{\overline{q^\prime}}\gamma^\mu q\;.
\label{10}
\ee
On a similar footing, the $0^-$ light meson particle (e.g. the pion) is
described by the effective operator:
\be
\Phi=\frac{1}{f_\pi}\overline{q^\prime}
\spur{\stackrel{\leftrightarrow}{\partial}} \gamma_5 q~.
\ee

We note explicitly that in both cases the approximation is expected to
work better in the limit of zero mass light mesons.
\par\noindent 
2) In the $H$-meson rest frame the two constituent quarks
have total momentum  equal to zero, whereas the sum of their energies:
$E_Q+E_q=\sqrt{k^2+m_Q^2}+\sqrt{k^2+m_q^2}$ is different from $m_H$
because of the presence of the interaction potential; to achieve a
considerable simplification, we assume 4-momentum conservation at each
hadron-quark-antiquark vertex and at the current-quark vertex. At the
same time, to describe the off-shell effect, following the prescription
first suggested in Ref. \cite{ACCMM}, we assume that the heavy quark has a
running mass $m_Q(k)$ defined by the energy conservation equation: 
\bea
E_Q&+&E_q=m_H \label{12}\\
E_Q&=& \sqrt{k^2+m_Q^2(k)} \label{13}\\
E_q&=& \sqrt{k^2+m_q^2}~. \label{14}
\eea
From previous equations, imposing $m^2_Q(k)\geq 0$, we obtain the
kinematical constraint
\be
0\leq k\leq k_M=\frac{m_H^2-m^2_q}{2 m_H}~,\label{14bis}
\ee
i.e. $k_M=2.64$ GeV for the $B-B^*$ system. We also note that, because of
the shape of the wave function, the average value of the running mass
for the $B$  meson is $m_Q(k)_{ave}\simeq 4.6$, which is  only slightly
different from the mass used in the fit ($4.89$ GeV). We also observe
that the shape of the wave function introduces an asymmetry between $m_Q$
and $m_q$. In principle, we could use (\ref{12}) to define the light
quark mass as running mass and this would give, for the $B$ meson,
$m_q(k)_{ave}=78$ MeV, to be compared to the result of the fit $m_q=38$
MeV. However, in this case we would obtain that the maximum value of
$k$ is $k_M\simeq 370$ MeV and, as a consequence, the constituent quarks
would be forbidden, by kinematical constraints, to reach the most likely
value $(k)_{ave} \simeq  500-600$ MeV (see Fig. \ref{f:fig1}). For this
reason one has to  use Eq. (\ref{12}) to define $m_Q(k)$ as the running
mass. 
\par
Let us note explicitly that this procedure distinguishes between the
constituent quarks, belonging to the heavy meson, that are on shell (the
off-shell effects being taken into account by the running mass
mechanism) and the other quarks, that we assume are able to move almost
freely in the hadronic matter and will be therefore described by the
free quark propagator modulated by a smooth shape function to take into
account off-shell effects. 
\par
\begin{figure}[ht]
\begin{center}
\input FEYNMAN
\begin{picture}(15000,15000)
\THICKLINES
\drawline\fermion[\E\REG](0,0)[5000]
\drawline\fermion[\W\REG](14900,0)[9900]
\drawarrow[\LDIR\ATTIP](\pmidx,\pmidy)
\put(8000,-2000){ $-q_2$}
\put(5000,3000){ $q_1$}
\put(13500,3000){ $p^\prime-q_2$}
\drawline\fermion[\E\REG](0,400)[5000]
\put(-5000,0){$H(p)$}
\put(5000,200){\circle*{700}}
\drawline\fermion[\NE\REG](5000,400)[7000]
\drawarrow[\LDIR\ATTIP](\pmidx,\pmidy)
\THINLINES
\drawline\fermion[\SE\REG](\pbackx,\pbacky)[7000]
\drawarrow[\LDIR\ATTIP](\pmidx,\pmidy)
\drawline\photon[\N\REG](\pfrontx,\pfronty)[4]
\put(14900,200){\circle*{700}}
\put(11000,8000){ $J_\mu$}
\drawline\fermion[\E\REG](14900,0)[5000]
\drawline\fermion[\E\REG](14900,400)[5000]
\put(21900,0){ $V(p^\prime)$}
\end{picture}
\vskip 0.3 cm
\caption{Heavy lines represent constituent quarks; 
the light line is the
(almost) free quark; $H=B,B^*$, $V=\rho,K^*$, $J_\mu$ is the current
inducing the decay.}
\label{f:fig2} 
\end{center}
\end{figure}
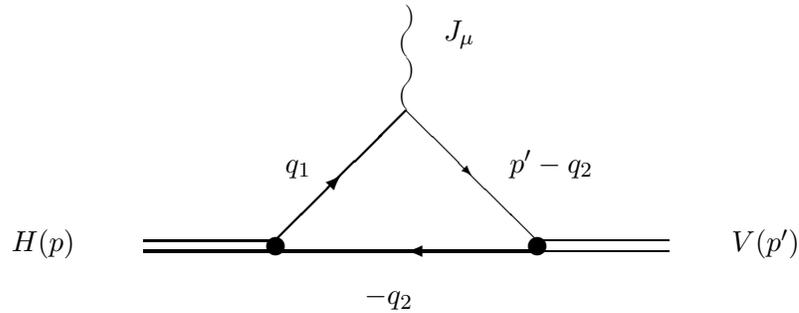

Let us now write down explicitly a set of  rules for the computation of
the hadronic matrix elements which implement these ideas. In order to
compute a typical matrix element such as (\ref{8}), a diagram like 
Fig. \ref{f:fig2} can be depicted with the following correspondences.

\par\noindent
1) For the heavy meson  $H$ in the initial state one introduces the
matrix:
\be
H=\frac{1}{\sqrt3}\psi (k){\sqrt{ \frac{m_q m_Q}
{m_q m_Q+q_1\cdot q_2} }}\;\;\frac{\spur{q_1}+m_Q}{2 m_Q}\Gamma 
\frac{-\!\!\!\spur{q_2}+m_q}{2 m_q}
\label{15}
\ee
where $m_Q=m_Q(k)$ is given by (\ref{12}), $1/\sqrt 3$ is a colour
factor, $q^\mu_1=(E_Q,\vec k),~q^\mu_2=(E_q,-\vec k)$ are  the
constituent quarks momenta, with $p^\mu=q^\mu_1+q^\mu_2$ ($p^\mu$ the
heavy meson momentum). $\Gamma$ is a matrix which is equal to
$-i\gamma_5$ for $J^P=0^-$ and $\spur \epsilon$ for $J^P=1^-$. We
observe that the factor ${\dd\sqrt{\frac{m_q m_Q} {m_q m_Q+q_1\cdot q_2}
}}$ has been introduced to enforce the normalization condition 
\be
<H|H>=2m_H
\ee
corresponding to Eq.(\ref{norm}). Finally, the wave function is given by
(\ref{wf}).
\par\noindent
2) For the heavy meson  $H$ in the final state the matrix:
\be
-\gamma^0 H^\dagger \gamma^0~. 
\label{16}
\ee
\par\noindent
3) For each quark line, not representative of a constituent quark, a
factor
\be
\frac{i}{\spur q-m_{q^{\prime}}}\times G(q^2)~.
\label{ff}
\ee
\par\noindent
As discussed above, this quark propagates almost freely in the hadronic
matter. For the shape function $G(q^2)$ we assume 
\be
G(q^2)=\frac{m^2_G-m^2_{q^{\prime}}} {m^2_G-q^2}
\label{ff2}
\ee
\par\noindent 
where $m_G$ is a parameter to be fitted.
Eq.(\ref{ff2}) corresponds to the Pauli-Villars regularization of the quark 
propagator, with mass $m_G$.
\par\noindent
4) For a light vector meson of polarization vector $\epsilon$ and quark
content $q\prime,~\overline q$
in the initial state the matrix
\be
 N_{q}N_{q^\prime}\frac{m^2_V}{f_V} \spur\epsilon ~,\label{17}
\ee
where $f_{\rho}=0.152$ GeV$^2$,  $f_{K^*}=0.201$ GeV$^2$\cite{PDG}. The
factor
$N_q$ is given by:
\bea
N_q&=&\sqrt{\frac{m_q}{E_q}}~~~~{\rm(if ~q=constituent~ quark)}\nn \\
&=&~~1~~~~~~~{\rm (otherwise)}~.\label{nq}
\eea
The reason for the factor $N_q$ is due to a
different normalization between the constituent and the (almost) free
quarks.
\par\noindent
5) For a light pseudoscalar  meson $M$ of quark content
$q\prime,~\overline q$, the matrix
\be
N_{q}N_{q^\prime}\frac{1}{f_M}(\spur\ell - \spur\ell^\prime)\gamma_5
\label{17bis}
\ee
where $\ell^\mu,~\ell^{\prime \mu}$ are the quark momenta,
$f_M=f_\pi\simeq 130$ MeV
for pions and $f_M=f_K \simeq 160$ MeV for kaons.\par\noindent
5) For the hadronic current in Eq. (\ref{9}) the factor
\be
N_{q} N_{q^\prime}\Gamma^\mu\;.
\label{j}
\ee
\par\noindent
7) For each quark loop, a colour factor $N_c=3$, a trace over Dirac
matrices and an integration over $k$:
\be
\int\frac{d^3k}{(2\pi)^3}\theta[k_M-k]~,
\ee
where $\theta(x)$ is the Heaviside function implementing the Eq. (\ref{12}).

\section{Leptonic decay constant}
\label{s:4}

To compute the leptonic decay constant $f_B$ we assume the previous
rules and we immediately get, from Eq. (\ref{fb}), 
\be
f_B p^\mu=-\sqrt{3}\int\frac{d^3k}{(2\pi)^3}\theta[k_M-k]\psi
(k)\sqrt{\frac{m_q m_Q}
{m_q m_Q+q_1\cdot q_2}}Tr\left [\frac{\spur{q_1}+m_Q}{2
m_Q}\gamma_5\frac{-\spur{q_2}+
m_q}{2 m_q}N_q N_Q \gamma^\mu\gamma^5 \right ]~.
\ee
Working out this expression in the meson rest frame we obtain
\be
f_B=\frac{\sqrt{3}}{2 \pi^2 m_B}\int_0^{k_M} dk~k^2\psi (k)\frac{m_q
E_Q+ m_Q E_q}
{\sqrt{E_q E_Q(m_q m_Q+q_1\cdot q_2)}}\label{fh}
\ee
\par
Eq. (\ref{fh}) agrees, in the limit $m_Q\to \infty$, with the results
obtained in \cite{piet} by the same model, but without the introduction
of the running mass and the trace formalism. To prove the formal
equivalence of the two approaches we perform the heavy quark limit in
(\ref{fh}): $m_Q(k)\simeq m_Q(k)_{ave}\simeq m_B\gg k,m_q$,
obtaining 
\be
f_B=\frac{\sqrt{3}}{4 \pi^2 m_B}\int dk~k^2\psi (k)\sqrt{\frac{m_q
+E_q}{E_q}}\lq
1-\frac{E_q-m_q}{2 m_B}\rq~,\label{fhold}
\ee
which agrees with result of ~\cite{piet} in the same limit. Numerically,
and for finite mass, the results of (\ref{fh}) and Ref.~\cite{piet}
differ, for the $B$ meson, by $10\%$, which gives an estimate of the
theoretical uncertainties of this procedure for the $B$ system. In the
charm case the deviations are higher (of the order $30-40\%$). This shows
that, to apply this formalism to the $D-D^*$ system, 
finite heavy quark mass effects must be 
properly taken into account. 

\section{ $B \to V$ form factors}
\label{s:5}

Let us now apply the previous formalism to the study of the form factors
describing the semileptonic decays $B\to\rho \ell\nu$ and $B\to
K^*\gamma$,  $B \to K^* \ell^+ \ell^-$. 
The corresponding  matrix elements can be written as 
follows:
\bea
<V(\epsilon(\lambda),p^\prime)|\overline{q^\prime}\gamma_\mu(1-\gamma_5)Q|B(p)>
& = & \frac{2 V(q^2)}{m_B+m_V}\epsilon_{\mu\nu\alpha\beta}
\epsilon^{*\nu}p^\alpha p^{\prime\beta}\nn\\
&-& i \epsilon^{*}_{\mu}(m_{B} + m_{V})  A_{1}(q^{2})
\nn\\
&+& i (\epsilon^{*}\cdot q) 
\frac{(p + p^\prime)_{\mu}}{m_B +  m_V}  A_{2}(q^{2})
\nn\\
&+& i  (\epsilon^{*}\cdot  q) 
\frac{2  m_V}{q^{2}} q_{\mu} [A_{3}(q^{2})  - A_{0}(q^{2})]
\;\; ,
\label{23}
\eea
where
\be
A_{3}(q^{2})  = \frac{m_{B} + m_{V}}{2  m_{V}} A_{1}(q^{2})
- \frac{m_{B} - m_{V}}{2  m_{V}} A_{2}(q^{2}) \;\; ,
\label{24}
\ee
and
\bea
<V(\epsilon(\lambda),p^\prime)|\overline{q^\prime}
\sigma_{\mu\nu}q^{\nu}{(1+\gamma_5)\over 2}Q|B(p)>
& = &
2 T_1(q^2)i\epsilon_{\mu\nu\alpha\beta}
\epsilon^{*\nu}p^\alpha p^{\prime\beta}\nn\\
&+& T_2(q^2)\lq  \epsilon^{*}_{\mu}(m^2_B -m_{V}^2)-(\epsilon^* \cdot
p)(p+p^\prime)_\mu\rq
\nn\\
&+&
T_3(q^2) (\epsilon^* \cdot p)\lq
q_\mu-\frac{q^2}{m^2_B-m^2_V}(p+p^\prime)_\mu\rq~.
\eea

At $q^2=0$ the following conditions hold
\bea
A_{3}(0)  &=& A_{0}(0)\nn\\
T_1(0)&=&T_2(0)\;.
\label{25}
\eea
\par
Let us write explicitly the matrix element of the tensor current:
\bea
<V(\epsilon(\lambda),p^\prime)|
\overline{q^\prime}\sigma_{\mu\nu}q^{\nu}(1 &+& \gamma_5)
Q|B(p)> = \nn\\
&=&\frac{N_Q N_q m_V^2}{\sqrt 3 f_V}\int\frac{d^3k}{(2\pi)^3}
\theta[k_M-k]\psi(k)
{\sqrt { \frac{m_q m_Q}
{m_q m_Q+q_1 \cdot q_2} }}G\lq (q_1-q)^2\rq \nn\\
&&Tr \Bigg[ \frac{\spur{q_1}+m_Q}{2 m_Q}\gamma_5
\frac{-\spur{q_2}+m_q}{2 m_q}
\spur{\epsilon}^*\frac{\spur{q_1}-
\spur{q}+m_{q^\prime}}{(q_1-q)^2-m^2_{q^\prime}}
\sigma_{\mu\nu}q^{\nu}(1+\gamma_5) \Bigg]\;.
\label{15b}
\eea
In a similar way we write all the other matrix elements. Working out the
trace and performing the angular integrations we get the analytic
formulae for the form factors reported in the Appendix. All
these form factors depend on the shape function $G$ defined in
(\ref{ff2}); in order to fix the unknown mass parameter $m_G$ we
consider the ratio 
\be
\frac{\Gamma (B \to K^*\gamma)}{\Gamma(b \to s
\gamma)}=4\Bigg(\frac{m_B}{m_b}\Bigg)^3
\Bigg(1-\frac{m^2_{K^*}}{m^2_B}\Bigg)^2|T_1(0)|^2~.
\ee
From the experimental results (\ref{cleorari}) we obtain
\be
T_1(0)=0.19 \pm 0.05~,
\ee
where we have used $m_b=4.8$ GeV.
Using for $T_1(0)$ the result given in the Appendix, we obtain
\be
m_G^2 \approx 3~{\rm GeV^2}~.
\ee
We can now compute, using the formulae given in the Appendix all the
form factors. Their values at $q^2=0$ are as follows. 
For $B\to\rho$: 
\be
\begin{array}{ccccccc}
V(0)   & = & 0.45~\pm~0.11  &~~~& A_2(0) & = & 0.26~\pm~0.05 \\
A_1(0) & = & 0.27~\pm~0.06  &~~~& A_0(0) & = & 0.29~\pm~0.09~.
\end{array}
\label{r1}
\ee
For $B\to K^*$
\be
\begin{array}{ccc}
T_1(0)=T_2(0) & = & 0.19~\pm~0.05\\
T_3(0)        & = & 0.43~\pm~0.08\;.
\end{array}
\label{r2}
\ee
These errors are obtained by varying $m_G^2$ in the range $1.3 \div
7.6$ $GeV^2$ corresponding to the errors in Eq. (\ref{cleorari}). In passing we
observe that our results depend smoothly on the mass parameter $m_G$. 

\begin{figure}[ht]
\begin{center}
\begin{tabular}{cc}
\epsfig{file=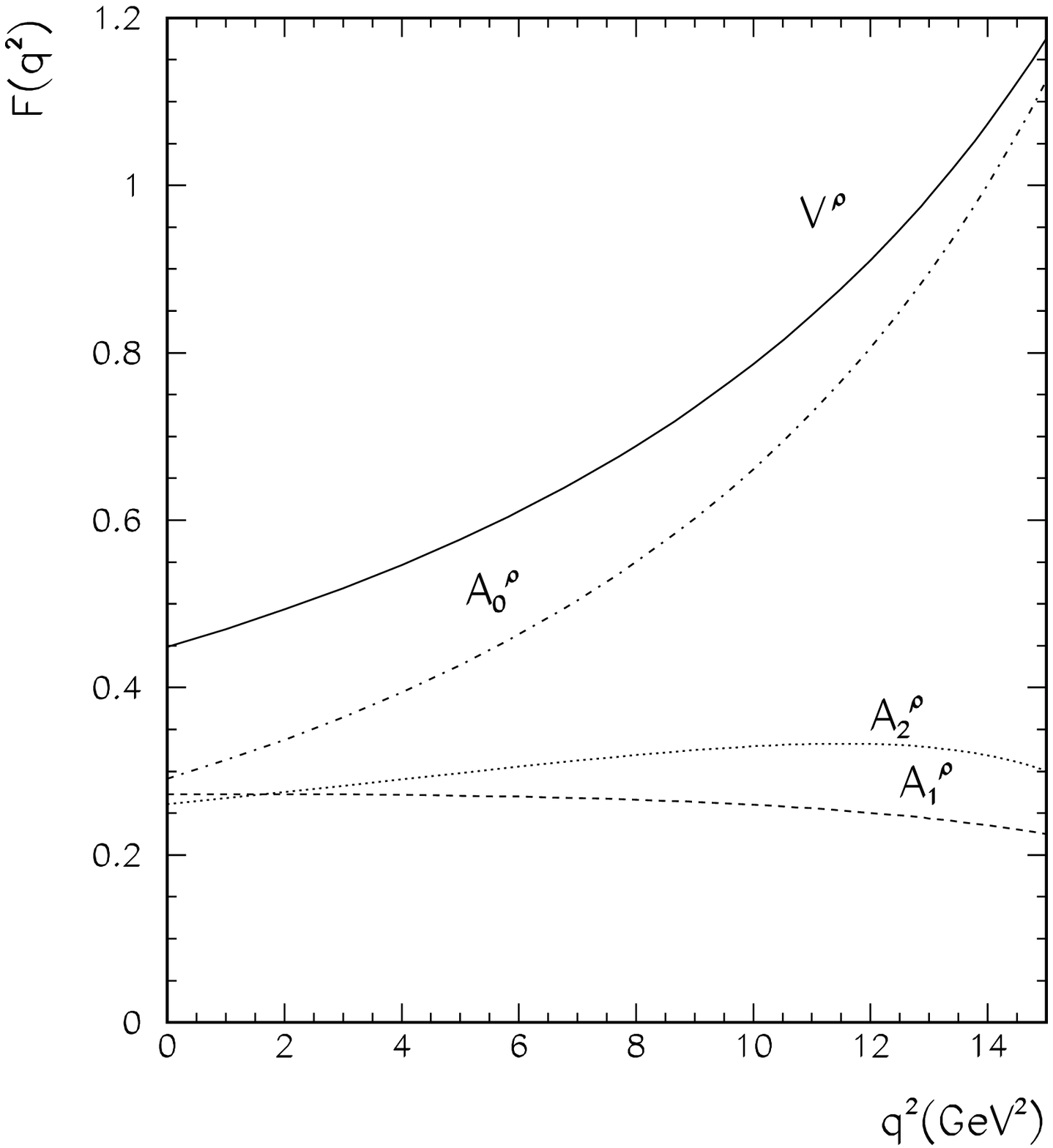,height=8cm} & 
\epsfig{file=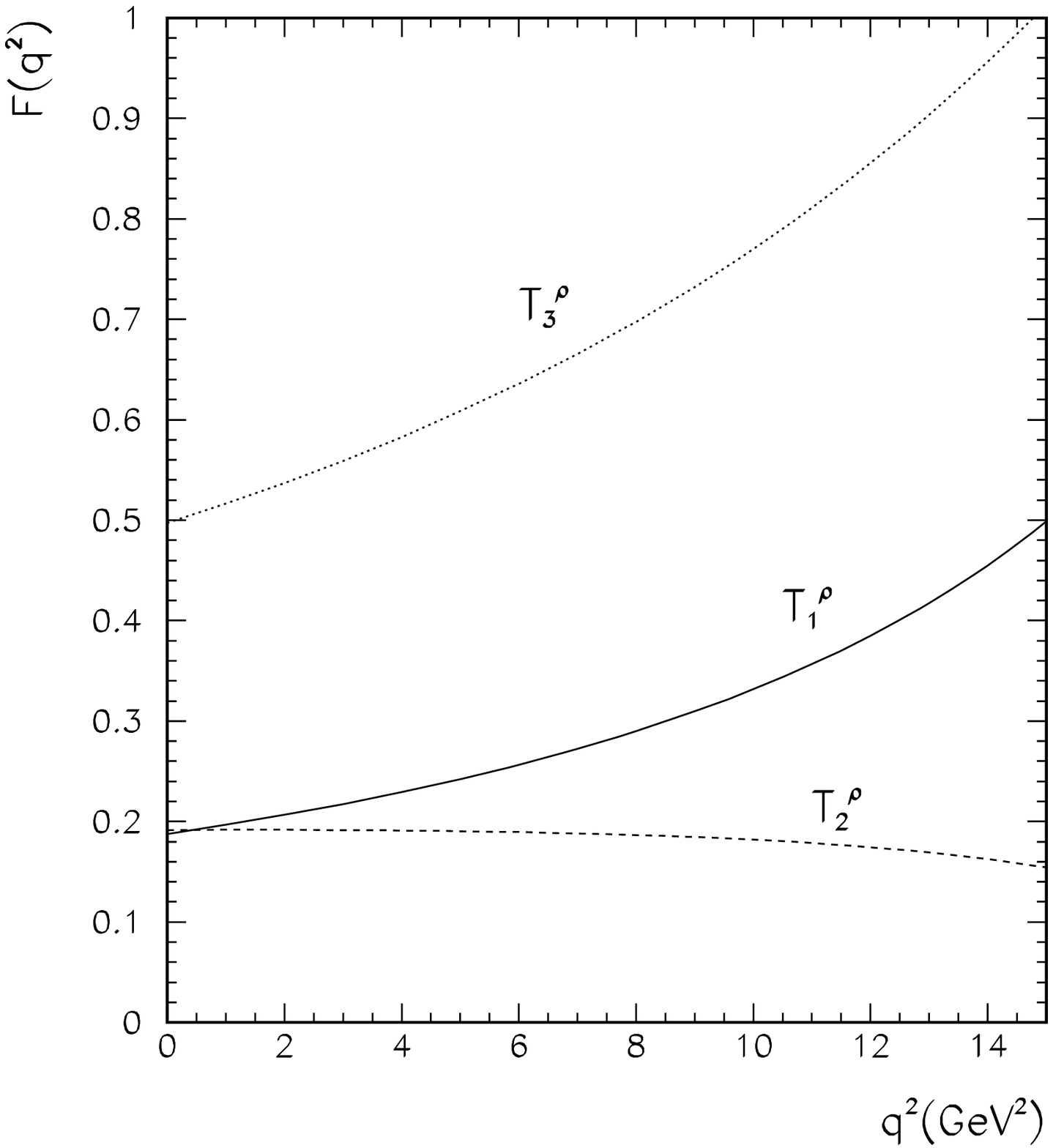,height=8cm} \\
\epsfig{file=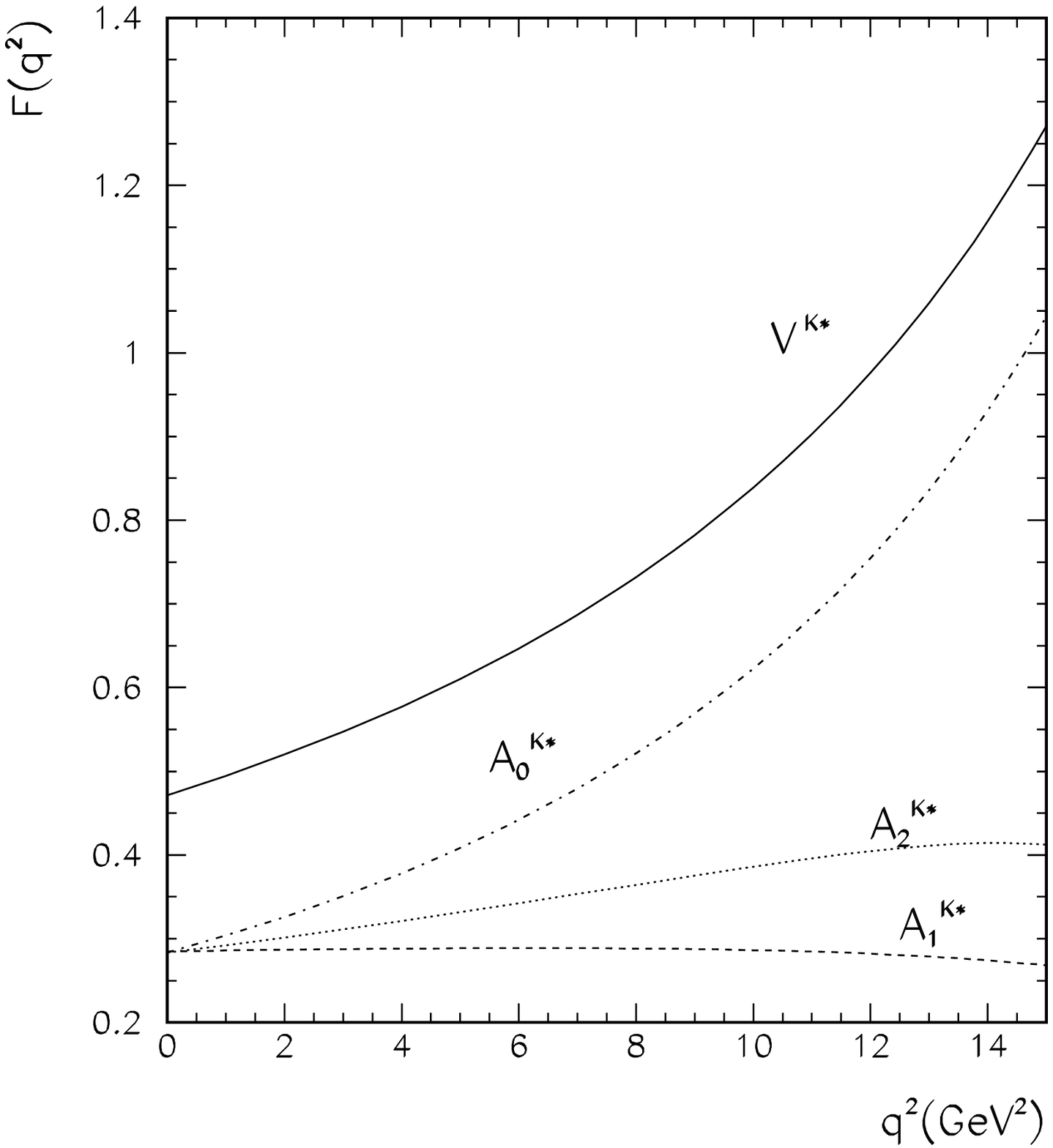,height=8cm} & 
\epsfig{file=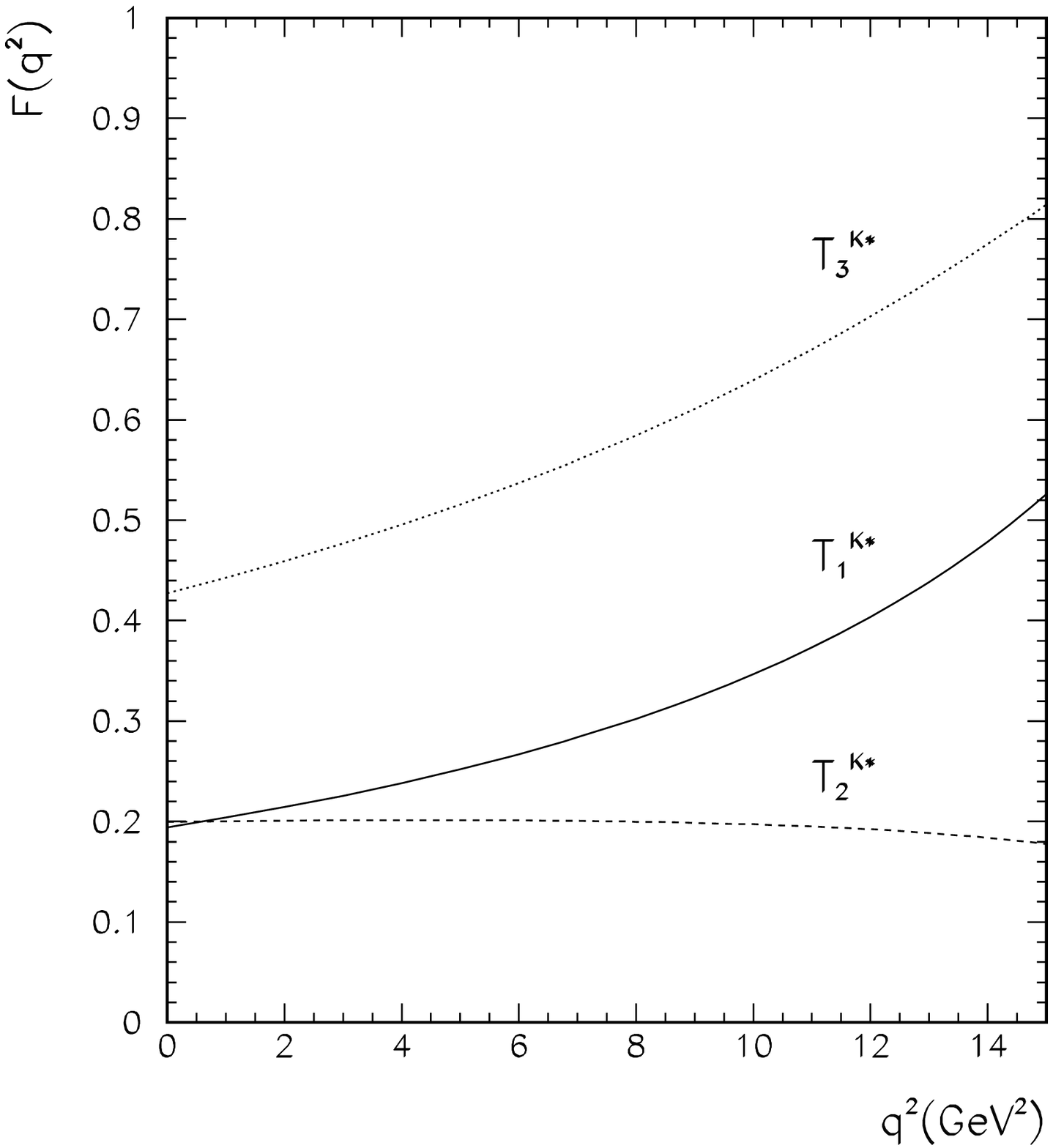,height=8cm} 
\end{tabular}
\end{center}
\caption{$q^2$ behaviour of  semileptonic and rare  B form 
factors. From left to right and from up to down: $B \to \rho$ (semileptonic),
$B \to \rho$ (rare), $B \to K^*$ (semileptonic), $B \to K^*$ (rare).}
\label{f:fig3} 
\end{figure}

In Fig. \ref{f:fig3} 
we report the $q^2$ dependence of the form factors $V,~A_1,~A_2,~A_0$ 
$T_1,~T_2$ and $T_3$ for the transitions $B\to \rho$ and $B\to K^*$.

\section{Comparison with the data and other theoretical approaches}
\label{s:6}

In order to give predictions on partial decay widths, we choose to fit
our theoretical results for the form factors by introducing the following
parameterization:
\be
F(q^2)=\frac{F(0)}{1~-~a_F \left(\dd\frac{q^2}{m_B^2}\right) +~b_F 
\left(\dd\frac{q^2}{m_B^2}\right)^2}
\label{16b}
\ee
\noindent
where $a_F~,b_F~$ are parameters to be fitted by  means of the 
numerical analysis performed up to $q^2=15~GeV^2$, both for $\rho$
and $K^*$ mesons. We collect the fitted values in Table \ref{t:tab1}.

\begin{table}[ht!]
\begin{center}
\begin{tabular}{||c||c|c|c||c|c|c|c||}
\hline\hline
& $F(0)$ & $a_F$ & $b_F$ & $F(0)$ & $a_F$ & $b_F$ & \\
\hline \hline 
$V^{\rho}$  & 
$0.45$ & $1.3$ & $0.27$ & 
$0.47$ & $1.3$ & $0.28$ & 
$V^{K^*}$\\ 
$A_0^{\rho}$ & 
$0.29$ & $1.9$ & $1.0$  &  
$0.28$ & $1.9$ & $0.94$  & 
$A_0^{K^*}$\\
$A_1^{\rho}$ & 
$0.27$ & $0.18$ & $0.96$  &  
$0.28$ & $0.19$  & $0.52$  & 
$A_1^{K^*}$\\
$A_2^{\rho}$ & 
$0.26$ & $1.0$ & $1.3$  &  
$0.28$ & $0.99$ & $0.71$  & 
$A_2^{K^*}$\\
$T_1^{\rho}$ & 
$0.19$ & $1.3$ & $0.29$ &  
$0.19$ & $1.3$ & $0.29$ & 
$T_1^{K^*}$\\
$T_2^{\rho}$ & 
$0.19$ & $0.21$ & $1.1$   &  
$0.19$ & $0.25$ & $0.80$  & 
$T_2^{K^*}$\\
$T_3^{\rho}$ & 
$0.50$ & $1.1$ & $0.22$ &  
$0.43$ & $0.99$ & $0.19$ & 
$T_3^{K^*}$\\
\hline \hline
\end{tabular}
\caption{Parameters of the various $B$ form factors.}
\label{t:tab1}
\end{center}
\end{table}

\par
From the table and from fig. \ref{f:fig3} one can see that $V(q^2)$,$T_1(q^2)$, 
$T_3(q^2)$ and $A_0(q^2)$ have a $q^2$ 
behaviour similar to a single pole; on the 
other hand, the other form factors have a practically flat  behaviour. In 
particular, $A_1(q^2)$ shows a slight decrease. A similar behaviour is obtained 
by 3-point sum rules \cite{ballnoi}, but not by the light cone sum rules
\cite{ballbraun}. 

 In Table 2 we compare our outcome for the values at $q^2=0$ with the results 
of other theoretical approaches.

\begin{table}[ht!]
\begin{center}
\begin{tabular}{||c||c|c|c|c||c|c||}
\hline \hline
& $This~work$ & 
$LCSR$ \cite{ballbraun} & 
$LCSR$ \cite{alibraun} & 
$LCSR$ \cite{aliev} & 
$SR$ \cite{ballnoi} & 
$\begin{array}{c}
Latt.~+ \\
LCSR \cite{lattice}
\end{array}$\\
\hline \hline 
$V^{\rho}(0)$   & $0.45~\pm~0.11$  & 
$0.34~\pm~0.05$  &
$0.35~\pm~0.07$  & $0.37 \pm 0.07$ &
$0.6~\pm~0.2$  & 
$0.35^{+0.06}_{-0.05}$ \\
$A_0^{\rho}(0)$ & $0.29~\pm~0.09$  & $ $&  & $    $  & $0.24 \pm 0.02   $  & 
$0.30^{+0.06}_{-0.04}$ \\
$A_1^{\rho}(0)$ & $0.27~\pm~0.06$  & 
$0.26~\pm~0.04$ &
$0.27~\pm~0.05$  & $0.30 \pm 0.05$ &
$0.5~\pm~0.1$  & 
$0.27^{+0.05}_{-0.04}$ \\
$A_2^{\rho}(0)$ & $0.26~\pm~0.05$  & 
$0.22~\pm~0.03$ &
$0.28~\pm~0.05$  & $0.33 \pm 0.05$ &
$0.4~\pm~0.2$  & 
$0.26^{+0.05}_{-0.03}$ \\
$T_1^{\rho}(0)$ & $0.19~\pm~0.05$  & 
$0.15~\pm~0.02$ &
$0.12~\pm~0.04$  & $0.15 \pm 0.05$ &
$   $  & $    $ \\
$T_3^{\rho}(0)$ & $0.50~\pm~0.08$  & 
$0.10~\pm~0.02$ &
$    $  &$ 0.10 \pm 0.05$     &
$   $  & $    $ \\
\hline \hline
$V^{K^*}(0)$   & $0.47~\pm~0.11$  & 
$0.46~\pm~0.07$ & 
$0.38~\pm~0.08$  &$0.45 \pm 0.08$
& $0.47~\pm~0.03$  & 
$    $ \\
$A_0^{K^*}(0)$ & $0.28~\pm~0.09$  & 
$ $ &
$    $  &  &$0.30~\pm~0.03$  & $    $ \\
$A_1^{K^*}(0)$ & $0.28~\pm~0.07$  & 
$0.34~\pm~0.05$ &
$0.32~\pm~0.06$  & $0.36 \pm 0.05$ &
$0.37~\pm~0.03$  & 
$ $ \\
$A_2^{K^*}(0)$ & $0.28~\pm~0.05$  & 
$0.28~\pm~0.04$ &
$$  & $0.40 \pm 0.05 $&
$0.40~\pm~0.03$  & $    $ \\
$T_1^{K^*}(0)$ & $0.19~\pm~0.05$  & 
$0.19~\pm~0.03$ &
$0.16~\pm~0.03$  & $0.17 \pm 0.05$&
$0.19~\pm~0.03$  & 
$0.16^{+0.02}_{-0.01}$ \\
$T_3^{K^*}(0)$ & $0.43~\pm~0.08$  & 
$0.13~\pm~0.02$ &
$    $  &  $0.13 \pm 0.05$ &$0.3$  & $    $ \\
\hline \hline
\end{tabular}
\caption{Comparison of the results coming from different works
on form factors.}
\label{t:tab2}
\end{center}
\end{table}

We also report the predictions for the
branching ratio ${\cal B}({\bar B}^0 \to \rho^+ \ell \nu)$ 
and for the partial widths at fixed helicity:

\be
{\cal B}(\bar B^0 \to \rho^+ \ell \nu) = 2.4 \times 10^{-4} \label{br}
\ee

\be
\begin{array}{c}
\Gamma_0 = 2.4 \times 10^{-17} \; s^{-1} \\
\Gamma_+ = 4.6 \times 10^{-18} \; s^{-1} \\
\Gamma_- = 7.4 \times 10^{-17} \; s^{-1} 
\end{array}
\ee
where $\Gamma_0$, $\Gamma_+$, $\Gamma_-$ refer to the $\rho$ helicities.
One can see that there is 
agreement between the result (\ref{br}) and the experimetal data in 
eq. (\ref{cleorho}).

In conclusion, the calculation based on the present QCD relativistic quark 
model seems quite adequate to describe the weak transition $B \to$ light vector 
meson. In spite of its simplicity the model embodies many of the features of 
more fundamental approaches; in particular 
it is confining and it contains the perturbative 
QCD $\alpha_s$ corrections through the Coulombic behaviour of the potential at 
small distances. Therefore it can be seen as a rather realistic model of the 
fundamental QCD description of these important weak processes.

\newpage
\appendix
\section{Form factors}

In this Appendix we report the expressions of the various 
form factors for the weak transitions $B(p) \to V(p^\prime, \epsilon) 
\ell \nu$. We note that, according to the discussion after eq. \ref{correl}, 
we put $p^{\prime 2}=0$.
\bea
V(q^2,x)&=&{m_B+m_V \over m_B}\;{m_V^2 \over f_V}{\sqrt{3} \over 8
\pi^2}
\int_0^{k_M} {dk\; u(k) \over  \sqrt{E_Q E_q(m_Q m_q+E_QE_q+k^2)}}
{1 \over \Big(1-{q^2\over m_B^2}\Big)}  \nonumber \\
& &\Bigg\{{2k \over \qv}\;{(m_Q-m_{q^\prime})m_B-(m_Q-m_q)q^0 \over m_B}
\label{v}\\
&+&\Bigg[ {m_q E_Q+m_Q E_q \over m_B}-{(m_Q-m_{q^\prime})m_B-(m_Q-m_q)q^0
\over m_B}
{2\qv  E_q-m_q^2+x^2 \over 2 \qv^2}\Bigg] \ln g(q^2,k,x)
\Bigg\} \nonumber
\eea
\par
\bea
A_1(q^2,x)&=&{1 \over (m_B+m_V) m_B}\;{m_V^2 \over f_V}{\sqrt{3} \over 4
\pi^2}
\int_0^{k_M} {dk\; u(k) \over \sqrt{E_Q E_q(m_Q m_q+E_QE_q+k^2)}}
{1 \over \Big(1-{q^2\over m_B^2}\Big)} \nonumber \\
&&\Bigg\{-(m_Q-m_q)\;{k \over \qv}\;(2\qv E_q -m_q^2+x^2)+ 2(m_Q-m_q)k
\;\qv  \label{a1}\\
&+& \Bigg[-q^0(m_Q E_q+m_q E_Q)+(m_{q^\prime}+m_Q)(m_B E_Q-m_Q^2+m_Q
m_q)\nonumber \\
&-&
(m_Q-m_q){2\qv E_q -m_q^2+x^2 \over 2}
+ k^2(m_Q-m_q)\Bigg( {(2\qv E_q -m_q^2+x^2)^2 \over 4 \qv^2 k^2}-1
\Bigg) \Bigg] \; \ln g(q^2,k,x)
\Bigg\} \nonumber
\eea
\par
\bea
A_2(q^2,x)&=&-{ (m_B+m_V)\over  m_B}\;{m_V^2 \over f_V}{\sqrt{3} \over 4
\pi^2}
\int_0^{k_M} {dk\; u(k) \over \sqrt{E_Q E_q(m_Q m_q+E_QE_q+k^2)}}
{1 \over \Big(1-{q^2\over m_B^2}\Big)} \nonumber \\
&& \Bigg\{ {k \over \qv}\; \Bigg[ m_{q^\prime}+m_Q+{q^0 \over m_B}(m_q-3m_Q)+
{2(m_Q-m_q) E_Q q^2 \over m_B^3} \Bigg] \nonumber\\
&-&
{2\qv E_q -m_q^2+x^2 \over 2 m_B^2 \qv^2} (m_Q-m_q)(2\qv-3m_B)k
\label{a2} \\
&+& \Bigg[ {m_Q-m_q \over  2m_B \qv}k^2-
{E_Q-E_q \over 2 m_B^2} (m_q E_Q+m_Q E_q)
+{m_Q-m_q \over m_B^2}(2\qv-3m_B) {(2\qv E_q -m_q^2+x^2)^2 \over 8
\qv^3}
\nonumber \\
&-& {2\qv E_q -m_q^2+x^2 \over 4 \qv^2}\Bigg(m_{q^\prime}+m_Q +{q^0\over m_B}
(m_q-3m_Q)+{2(m_Q-m_q)E_Q q^2 \over m_B^3} \Bigg)\Bigg]\;
\ln g(q^2,k,x) \Bigg\} \nonumber
\eea
\par
\bea
A_0(q^2,x)&=&{m_V \over  f_V}{\sqrt{3} \over 8 \pi^2 m_B}
\int_0^{k_M} {dk\; u(k) \over \sqrt{E_Q E_q(m_Q m_q+E_QE_q+k^2)}}
{1 \over \Big(1-{q^2\over m_B^2}\Big)} \nonumber \\
&&\Bigg\{ -2k \;\qv \Bigg[ {2(m_Q-m_q)E_Q \over m_B^2 \qv}q^2 -{1 \over
m_B \qv}
\Bigg( (m_Q+m_{q^\prime})m_B q^0+(m_Q-m_q)q^2 \Bigg)-m_{q^\prime}+m_Q \Bigg] 
\nonumber \\
&+&{2k\;(m_Q-m_q) \over m_B}(2\qv E_q -m_q^2+x^2) \label{a0}\\
&+& \Bigg[m_Q(m_q-m_Q)(m_Q+m_{q^\prime})+m_Q q^2 -{m_Q-m_q \over m_B}
{(2\qv E_q -m_q^2+x^2)^2 \over 2\qv} \nonumber \\
&+& {E_Q \over m_B} \Bigg(-2m_B m_Q q^0+(m_Q-m_q)(2q^0 E_Q-q^2)+
m_B^2(m_Q+m_{q^\prime})\Bigg) \nonumber \\
&+&
\Bigg({2(m_Q-m_q)E_Q \over m_B^2 \qv}q^2 -{1 \over m_B \qv}
\Big( (m_Q+m_{q^\prime})m_B q^0+(m_Q-m_q)q^2 \Big)-m_{q^\prime}+m_Q  \Bigg) 
\Bigg]\;
\nonumber \\
&&{ 2\qv E_q -m_q^2+x^2 \over 2} \ln g(q^2,k,x)
\Bigg\}\;. \nonumber
\eea
\par\noindent
Here:
\be
\qv={m_B^2-q^2 \over 2} \hskip 1.5 cm q^0=\sqrt{q^2+\qv^2}
\ee
\be
g(q^2,k,x)={|2k\qv+2\qv E_q-m_q^2+x^2| \over |-2k\qv+2\qv E_q-m_q^2+x^2|}\;.
\ee

The results for the form factors describing the decay 
$B(p) \to V(p^\prime,\epsilon) \gamma$ are
as follows $(p^{\prime 2}=0)$: 

\bea
T_1(q^2,x) &=& {m_{V}^2 \over f_{V}} {\sqrt{3} \over 16 \pi^2 m_B}
\int_0^{k_M} {dk\; u(k) \over \sqrt{E_b E_q(m_b m_q+E_b E_q+k^2)}}
{1 \over \Big(1-{q^2\over m_B^2}\Big)} \nonumber \\
&&\Bigg\{-2 {k \over \qv}\Bigg[q^2-2 E_b q^0+{q^0 \over
m_B}(m_b+m_{q^\prime})(m_b-m_q)
\Bigg]-{k \over \qv}(2\qv E_q-m_q^2+x^2) \nonumber \\
&+& \Bigg[(m_b+m_{q^\prime}){m_b E_q+m_q E_b \over m_B}+k^2+
{(2\qv E_q-m_q^2+x^2)^2 \over 4 \qv^2} \label{t1} \\
&+&{2\qv E_q-m_q^2+x^2 \over 2} \Bigg( {(m_b+m_{q^\prime})(m_b-m_q)q^0 
\over m_B \qv^2}
+{q^2-2 E_b q^0 \over \qv^2} \Bigg) \Bigg] \;
\ln g(q^2,k,x)
\Bigg\} \nonumber
\eea
\par
\bea
T_2(q^2,x) &=& {1 \over m_B^2-m_V^2}
{m_{V}^2  \over f_{V}} {\sqrt{3} \over 8 \pi^2 m_B}
\int_0^{k_M} {dk\; u(k) \over \sqrt{E_b E_q(m_b m_q+E_b E_q+k^2)}}
{1 \over \Big(1-{q^2\over m_B^2}\Big)} \nonumber \\
& &\Bigg\{2k\; \qv [(m_b+m_{q^\prime})(m_q-m_b)+2 m_B E_b]-m_B q^0 k {2\qv
E_q-m_q^2+x^2
\over \qv} \nonumber  \\
&+& \Bigg[q^0[m_B m_b m_{q^\prime}+E_b(m_b+m_{q^\prime})(m_q-m_b)+m_B
E_b^2]+q^2[m_b(m_b-m_q)-m_B
E_b] \nonumber  \\
&-&[(m_b+m_{q^\prime})(m_q-m_b)+2 m_B E_b]{2\qv E_q-m_q^2+x^2 \over  2}\label{t2}
\\
&+&
m_B q^0 {(2\qv E_q-m_q^2+x^2)^2 \over 4 \qv^2}  \Bigg]
\ln g(q^2,k,x)
\Bigg\} \nonumber
\eea
\par
\bea
T_3(q^2,x) &=& {m_{V}^2 \over f_{V}}  {\sqrt{3} \over 8 \pi^2 m_B}
\int_0^{k_M}  {dk\; u(k) \over \sqrt{E_b E_q(m_b m_q+E_b E_q+k^2)}}
{1 \over \Big(1-{q^2\over m_B^2}\Big)} \nonumber \\
&&\Bigg\{ -k \Bigg[ -2E_b+{1 \over \qv}\Bigg(-m_B^2+{2 m_B-q^0 \over
m_B}
(m_b+m_{q^\prime})(m_b-m_q)\Bigg) \nonumber \\
&+& {2 \qv+3 q^0 \over 2 \qv^2}(2\qv E_q-m_q^2+x^2) \label{t3}
\Bigg]  \\
&+& \Bigg[{m_b(m_b-m_{q^\prime}) \over 2}+m_b(m_q-m_b)+m_B E_q+{E_b \over m_B}
{(m_b +m_{q^\prime})(m_b-m_q) \over 2} -{q^0 k^2 \over 2 \qv} \nonumber \\
&+& {2\qv E_q-m_q^2+x^2 \over 2 \qv} \Bigg( -E_b +{1 \over 2 \qv}
\Big(-m_B^2+{2 m_B -q^0 \over m_B}(m_b+m_{q^\prime})(m_b-m_q) \Big) \Bigg)
\nonumber\\
&+& {2 \qv+3 q^0 \over 8 \qv^3}(2\qv E_q-m_q^2+x^2)^2 \Bigg]\;
\ln g(q^2,k,x) \Bigg\}  \nonumber
\eea

\par
All the form factors are obtaind by taking the difference 
$F(q^2)=F(q^2,x=m_{q^\prime})-F(q^2,x=m_G)$, 
where $m_G$ is the mass parameter defined in the text; moreover 
$m_{q^\prime}=m_q$
when $V=\rho$, while $m_{q^\prime}=m_s$ when $V=K^*$.


\newpage

\begin{thebibliography}{9}

\bibitem{physbook}
For an updated review of the physics potential of $B$ factories see
"The BaBar Physics Book", SLAC-R-504 (in preparation).
%
\bibitem{cleo96}
CLEO Collab., J. P. Alexander et al., Phys. Rev. Lett. {\bf 77} (1996) 5000.
%
\bibitem{cleo95}
CLEO Collab., M. S. Alam et al., Phys. Rev. Lett. {\bf 74} (1995) 2885;
%
\bibitem{cleo93}
CLEO Collab., R. Ammar et al., Phys. Rev. Lett. {\bf 71} (1993) 674.
%
\bibitem{latticerev}
For review of lattice QCD results in $B$ physics see:
J.M. Flynn and C.T. Sachrajda, in "Heavy Flavours" (2nd ed.), 
ed. by A.J. Buras and M. Linder (World Scientific, Singapore).
%
\bibitem{sumrev}
For a review on QCD sum rules see: "Vacuum structure and QCD sum rules",
ed. M.A. Shifman (North Holland, Amsterdam, 1992).
%
\bibitem{salpeter}
E. E. Salpeter, Phys. Rev. {\bf 87} (1952) 328.
%
\bibitem{piet}
P. Cea, P. Colangelo, L. Cosmai and G. Nardulli,
Phys. Lett. {\bf B 206} (1988) 691; P. Colangelo, G. Nardulli and M.
Pietroni, Phys. Rev. {\bf D 43} (1991)  3002.
%
\bibitem{cea}P. Cea and G. Nardulli, Phys. Rev. {\bf D 34} (1986) 1863.
%
\bibitem{rich}J. L. Richardson, Phys. Lett. {\bf B 82} (1979) 272.
%
\bibitem{durand}L. Durand, Phys. Rev. {\bf D32} (1985) 1257.
%
\bibitem{multhopp}K. Karamcheti, "Principles of ideal fluid aerodynamics" 
(Wiley, New York, 1966).
%
\bibitem{tedesco}P. Colangelo, G. Nardulli and L. Tedesco, Phys. Lett.
{\bf B 272} (1991)
344.
%
\bibitem{thomas}V. Morenas, A. Le Yaouanc, L. Oliver, O. P\`ene, J.C.
Raynal, Phys. Lett. {\bf B 408} (1997) 357; Phys. Rev. {\bf D 56} (1997) 5668.
%
\bibitem{ACCMM}G. Altarelli, N. Cabibbo, G. Corb\`o, L. Maiani and G.
Martinelli, Nucl. Phys. {\bf B208} (1982) 365.
%
\bibitem{PDG}
C. Caso {\it et al.}, (Particle Data Group),  Eur. Phys. J. {\bf C 3} 
(1998) 1.
%
\bibitem{ballnoi}
P. Ball, Phys. Rev. {\bf D 48} (1993) 3190;
P. Colangelo, F. De Fazio, P. Santorelli, Phys. Rev. {\bf D 51} (1995) 2237;
P. Colangelo, F. De Fazio, P. Santorelli, E. Scrimieri,
Phys. Rev. {\bf D 53} (1996) 3672; {\bf D 57} (1998) 3186 (E).
%
\bibitem{ballbraun}
P. Ball and V. M. Braun, hep-ph/9805422.
%
\bibitem{alibraun}
A. Ali, V.M. Braun, and H. Simma, Z. Phys. {\bf C 63} (1994) 437;
P. Ball, and V.M. Braun, Phys. Rev. {\bf D 55} (1997) 5561.
%
\bibitem{aliev}T.M. Aliev, A. Ozpineci and M. Savci, Phys. Rev. {\bf  D 56} 
(1997) 4260.
%
\bibitem{lattice}
L. Del Debbio {\it et al.}, (UKQCD Collaboration),
Phys. Lett. {\bf B 416} (1998) 392.
\end{thebibliography}
\end{document}